\documentclass[12pt]{iopart}
\usepackage{graphicx}
\usepackage[dvips]{color}
\begin{document}

\title{Precision comparison of the quantum Hall effect in graphene and gallium arsenide}

\author{T.J.B.M. Janssen$^1$, J.M. Williams$^1$, N.E. Fletcher$^2$, R. Goebel$^2$, A. Tzalenchuk$^1$, R. Yakimova$^3$, S. Lara-Avila$^4$, S.~Kubatkin$^4$, and V.I.~Fal'ko$^5$}

\address{$^1$National Physical Laboratory, Hampton Road, Teddington TW11 0LW, UK}
\address{$^2$Bureau International des Poids et Mesures, Pavillon de Breteuil, F-92312, France}
\address{$^3$Department of Physics, Chemistry and Biology (IFM), Link\"{o}ping University, S-581 83 Link\"{o}ping, Sweden}
\address{$^4$Department of Microtechnology and Nanoscience, Chalmers University of Technology, S-412 96 G\"{o}tenborg, Sweden}
\address{$^5$Physics Department, Lancaster University, Lancaster LA1 4YB, UK}

\ead{jt.janssen@npl.co.uk}

\begin{abstract}
The half-integer quantum Hall effect in epitaxial graphene is compared with high precision to the well known integer effect in a GaAs/AlGaAs heterostructure. We find no difference between the quantised resistance values within the relative standard uncertainty of our measurement of $8.7\times 10^{-11}$. The result places new tighter limits on any possible correction terms to the simple relation $R_{\rm K}=h/e^2$, and also demonstrates that epitaxial graphene samples are suitable for application as electrical resistance standards of the highest metrological quality. We discuss the characterisation of the graphene sample used in this experiment and present the details of the cryogenic current comparator bridge and associated uncertainty budget. 
\end{abstract}

\maketitle

\section{Introduction}
The discovery 6 years ago of the quantum Hall effect (QHE) in graphene sparked an immediate interest in the metrological community~\cite{Novoselov2005,Zhang2005}. The QHE is a fascinating macroscopic quantum effect occurring in two-dimensional conductors and relates the resistance quantum, $h/e^2$ only to the fundamental constants of nature, $h$, the Planck constant and, $e$, the elementary charge~\cite{Prange1990}. Although the QHE has been used successfully in metrology to realise the resistance scale for more than two decades~\cite{Jeckelmann2001}, graphene is a material with properties like no other. Graphene, a single layer of carbon atoms in a hexagonal crystal lattice structure, is a truly two dimensional metal with a linear dispersion relationship characteristic of massless Dirac-type charge carriers~\cite{Geim2009}. The unique bandstructure of this semi-metal has both practical and fundamental implications. Firstly, the massless nature of the charge carriers leads to a Landau level spectrum with an energy gap between the first two levels which is around 5 times larger than that in semiconductor materials for magnetic fields around 10 tesla. This implies that the QHE in graphene can be observed at much reduced magnetic fields and/or much higher temperatures~\cite{Novoselov2007}. Secondly, the marked difference in bandstructure and charge carrier characteristics between graphene and semiconductor systems allows for a demonstration of the universality of the quantum Hall effect through a rigorous test of the material independence of the value of $R_{\rm K}$, the von Klitzing constant.  

Theory predicts no measurable corrections to the simple relation $R_{\rm K}=h/e^2$. The quantum Hall resistance is considered to be a topological invariant~\cite{Thouless1994}, not altered by the electron-electron interaction, spin-orbit coupling, or hyperfine interaction with nuclei. It has also been shown that the quantised Hall resistance is insensitive to much more subtle influences of the gravitational field~\cite{Hehl2004}. Recently, a quantum electrodynamical approach to charge carriers in a magnetic field has predicted a tiny correction to the von Klitzing constant of the order of $10^{-20}$ for practical magnetic field values~\cite{Penin2009}. However, the size of this predicted correction is about 10 to 8 orders of magnitude smaller than the most accurate measurement techniques available and therefore untestable. Nevertheless, the fundamental nature of the Hall resistance quantisation makes experimental tests of its universality of the utmost importance, in particular, for improving our knowledge of two fundamental quantities of nature: the electron charge and the Planck constant. The precision obtained through a universality test as presented here is much greater than is possible by a comparison of the best values of the constants $h$, $e$, and $R_{\rm K}$~\cite{Mohr2008b}. Analysis of the complete set of published results carried out by CODATA~\cite{Mohr2008b} showed no deviation from $h/e^2$ to within $2\times 10^{-8}$, which calls for more accurate measurements. Universality of $R_{\rm K}$ will strongly support the pending redefinition of the SI-units for mass and current in terms of $h$ and $e$~\cite{Mills2011}.

A direct comparison of the Hall resistance in two different substances does not prove the exactness of the relationship $R_{\rm K}=h/e^2$; however, material independence is a significant factor in establishing the fundamental nature of $R_{\rm K}$. This material independence turns out to be rather difficult to establish. Indeed the characteristics of QHE samples must satisfy very stringent requirements~\cite{Delahaye2003} and in 30 years only silicon MOSFETs (metal-oxide-semiconductor field-effect transistors) and III-V (GaAs/AlGaAs or InGaAs/InP) heterostuctures did so. 

The first accurate measurements of the QHE in graphene were performed by Giesbers {\it et al.}~\cite{Giesbers2008} on exfoliated samples. The precision obtained in these measurements was 15~ppm and limited by the high ($\approx\rm k\Omega$) contact resistances together with a small ($\approx\rm 2.5\ \mu A$) maximum source-drain current which these samples could sustain before breakdown of the QHE occurred. A large measurement current determines the maximum signal-to-noise ratio and increasing this breakdown current is key to high resolution measurements. One established method of increasing the breakdown current is to increase the sample width~\cite{Jeckelmann2001} which is not easy to achieve with the exfoliation technique. 

In 2009 a number of groups around the world almost simultaneously succeeded in growing large-area wafers of epitaxial graphene by sublimation of SiC  with a quality good enough to observe the quantised Hall resistance~\cite{Shen2009, Wu2009, Tzalenchuk2010, Jobst2010, Tanabe2010}. In an indirect comparison with a GaAs/AlGaAs device via an intermediate room-temperature standard resistor, we demonstrated universality of $R_{\rm K}$ with an uncertainty of 3~parts in $10^9$~\cite{Tzalenchuk2010}. The measurement system~\cite{Williams2010} was identical to that used by Giesbers {\it et al.}~\cite{Giesbers2008} and the key factors in the improvement were the very low contact resistances and large sample size which were achieved, resulting in an order of magnitude increase in the breakdown current. Subsequently we undertook a direct comparison between epitaxial graphene and two GaAs/AlGaAs devices using a modified measurement system demonstrating equivalence to an accuracy of 8.6 parts in $10^{11}$~\cite{Janssen2011b}. Recently, much progress has been made on metrological quantum Hall measurements in exfoliated graphene devices with accuracies continuously increasing to around 5 parts in $10^7$ \cite{Woszczyna2011,Guignard2011}.

In this paper we present a detailed description of the graphene device characterisation, measurement system, data analysis and resulting uncertainty budget which underpins the direct comparison result~\cite{Janssen2011b}.

\section{Characterisation of epitaxial graphene}\label{CharGraphene}

\begin{figure}
	\includegraphics{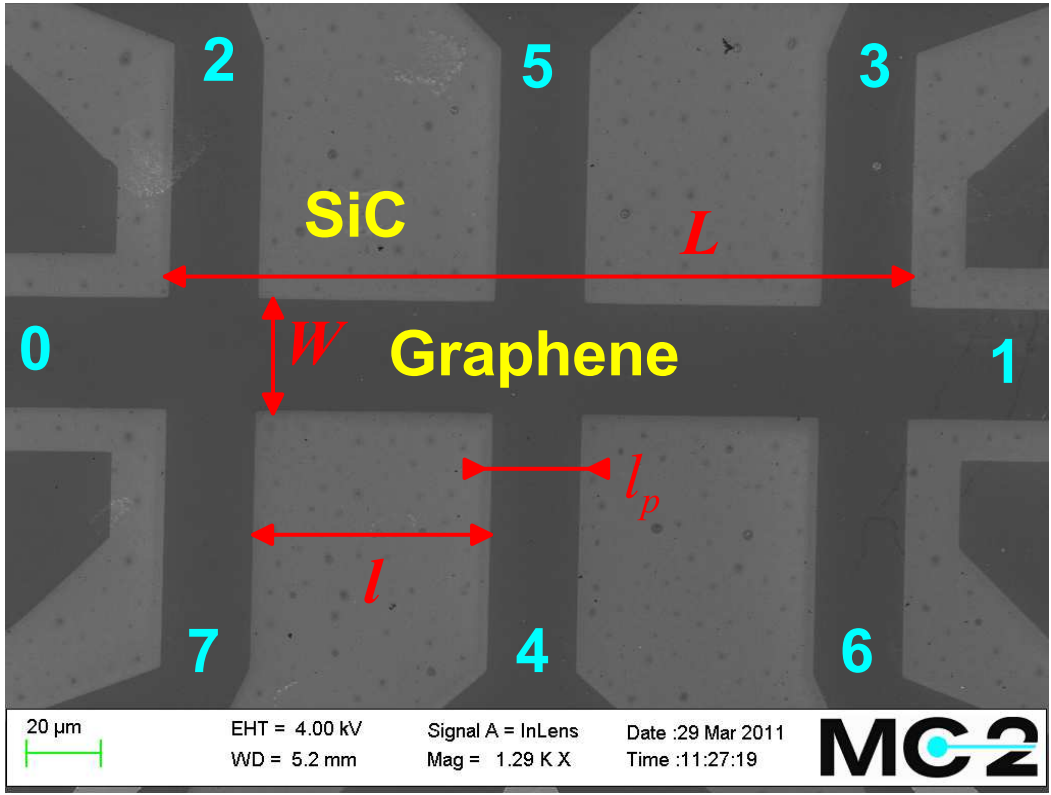}
	\caption{\label{fig1} SEM image of a typical device and contact labels. The values for the parameters $L$, $W$, $l$ and $l_p$ are listed in Tab.~\ref{tab2}.}
\end{figure}

In~\cite{Tzalenchuk2010} we reported the first accurate measurements of the 
quantum Hall effect in large high-quality epitaxial samples. The material studied 
in our experiments was grown on the Si-terminated face of a 4H-SiC(0001) substrate~\cite{Virojanadara2008}. The reaction kinetics on the Si-face are slower than on the C-face because of the higher surface
energy, which helps homogeneous and well-controlled graphene formation. Graphene
was grown at 2000~$^\circ$C and 1~atm Ar gas pressure, which result in monolayers of
graphene atomically uniform over more than 50~$\rm \mu m^2$, as shown by low-energy electron
microscopy. Twenty Hall bar devices of different sizes, from 160~$\rm \mu m\times 35\ \rm \mu m$ down to 
11.6~$\rm \mu m\times 2\ \rm \mu m$ were produced on each 0.5~$\rm cm^2$ wafer using standard electron beam
lithography and oxygen plasma etching (Figure \ref{fig1}). Atomic force microscopy (AFM)
images revealed that the graphene layer covers the substrate steps like a carpet,
preserving its structural integrity~\cite{Tzalenchuk2010}. Contacts to graphene were produced by
straightforward deposition of 3~nm of Ti and 100~nm of Au through a lithographically
defined mask followed by lift-off, with a typical area of graphene-metal interface of
$\rm 10^4\ \mu m^2$ for each contact. Using transport measurements in low magnetic fields, we established that the manufactured material was n-doped, with the measured electron concentration in the range of $\rm (5.5-9)\times 10^{11}\ cm^{-2}$, mobility about $\rm 2400\ cm^2V^{-1}s^{-1}$ at room temperature and between 4000 and $\rm 7500\ cm^2V^{-1}s^{-1}$ at 4.2~K, almost independent of device dimensions and orientation with respect to the substrate terraces. All results presented in this work were obtained on the largest Hall bar (i.e. 160~$\times$ 35~$\rm \mu m$).

In fact epitaxial graphene grown on SiC is always strongly n-type doped with carrier densities typically in the range of $10^{12}-10^{13}\ \rm cm^{-2}$. The doping of the graphene is caused by the so-called ``dead layer'' of carbon atoms in between the SiC substrate and graphene. This layer is non-conducting and characterised by a $6\sqrt{3}\times 6\sqrt{3}$ supercell of the reconstructed surface of sublimated SiC. Missing or substituted carbon atoms in various positions of such a huge supercell in the dead layer create localised surface states with a broad distribution of energies within the bandgap of SiC. It appears that the density of such defects is higher in material grown at low temperatures ($1200-1600\rm ^\circ C$) resulting in graphene  doped to a large electron density~\cite{Kopylov2010}. On the other hand, the material used in this work is grown at high temperatures in a highly pressurised atmosphere of Ar which seems to improve the integrity of the reconstructed 'dead' layer, leading to a lower density of donors on the surface and, therefore, producing graphene with a much lower initial doping~\cite{Tzalenchuk2010,Lara-Avila2011}.

\begin{figure}
\includegraphics{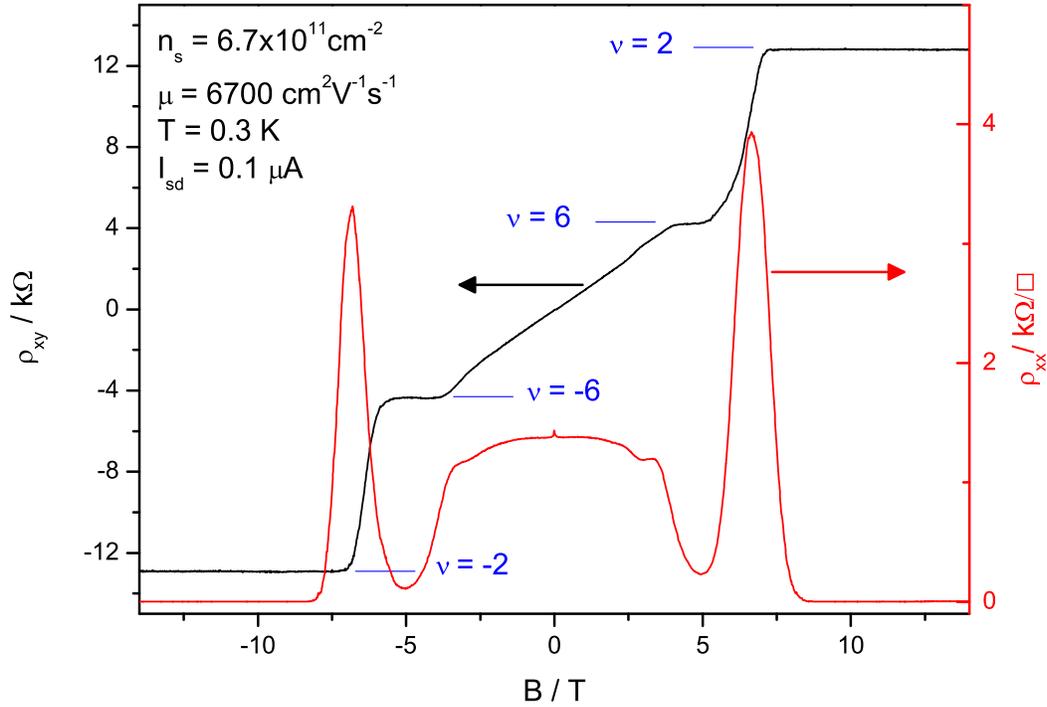}
\caption{\label{fig2} Transverse ($\rho_{xy}$) (contacts 4 and 5 in Fig.~\ref{fig1}) and longitudinal ($\rho_{xx}$) (contacts 6 and 7) resistivity measurement. In all measurements in this work the source-drain current is between contacts 0 and 1. The horizontal lines indicate the exact quantum Hall resistivity values for filling factors $\nu = \pm 2$ and $\pm 6$.}
\end{figure}

Figure~\ref{fig2} shows a typical measurement of the longitudinal and transverse resistivity as a function of magnetic flux density $B$ at low temperatures. Wide plateaux are observed in the transverse resistivity which are accompanied by a vanishingly small value of the longitudinal resistivity. The sequence of quantum Hall plateaux and absence of the $\nu =4$ plateau confirms that the sample is monolayer graphene. At zero field a small weak localisation peak is visible - a signature of quantum coherence in disordered Fermi liquid in epitaxial graphene~\cite{Lara-Avila2011b}. Charateristic scattering lengths have been determined from the analysis of the weak localisation correction to the Drude conductivity. For our graphene grown on SiC we find that the phase coherence length $L_\phi\approx 1\ \rm\mu m$ at low temperatures, the intervalley scattering length $L_i\approx 0.2\ \mu m$ and the intravalley scattering length $L_*\approx 0.02\ \rm\mu m$, the latter two being virtually temperature independent. These experimental scattering lengths set a limit of at least 50~ps on the spin relaxation time in this material.

For accurate quantum Hall measurements it is important that the contact resistance is well below $\approx 100\ \Omega$~\cite{Jeckelmann2001}. Table~\ref{Contacts} shows the measured contact resistances for our device and demonstrates that this condition is satisfied for most of the contacts. Our graphene device was approximately two years old at the time of these measurements and had been thermally cycled between 300~mK and room temperature more than a dozen times. Also the contacts have been re-bonded several times in the TO8-header which has caused damage to some of the bonding pads and is the most likely cause of the poor resistance of contacts $3$ and $5$ (on pristine devices we routinely get all contact resistances below 1~$\Omega$).

\begin{table}
\begin{center}
\caption{\label{Contacts} Three terminal contact resistances measured at $B=14\ \rm T$ in the $\nu=2$ quantum Hall state. The measurement current was $\rm 10\ \mu A$ and a lead resistance of 2.7~$\Omega$ was subtracted.}
\begin{tabular}{ll}
\br
Contact No.& $R(\Omega)$ \\
\mr
0 & 0.3 \\
1 & 0.3 \\
2 & 15  \\
3 & 125 \\
4 & 0.3 \\
5 & 90  \\
6 & 23  \\
7 & 0.3 \\
\br
\end{tabular}
\end{center}
\end{table}

\begin{figure}
\includegraphics{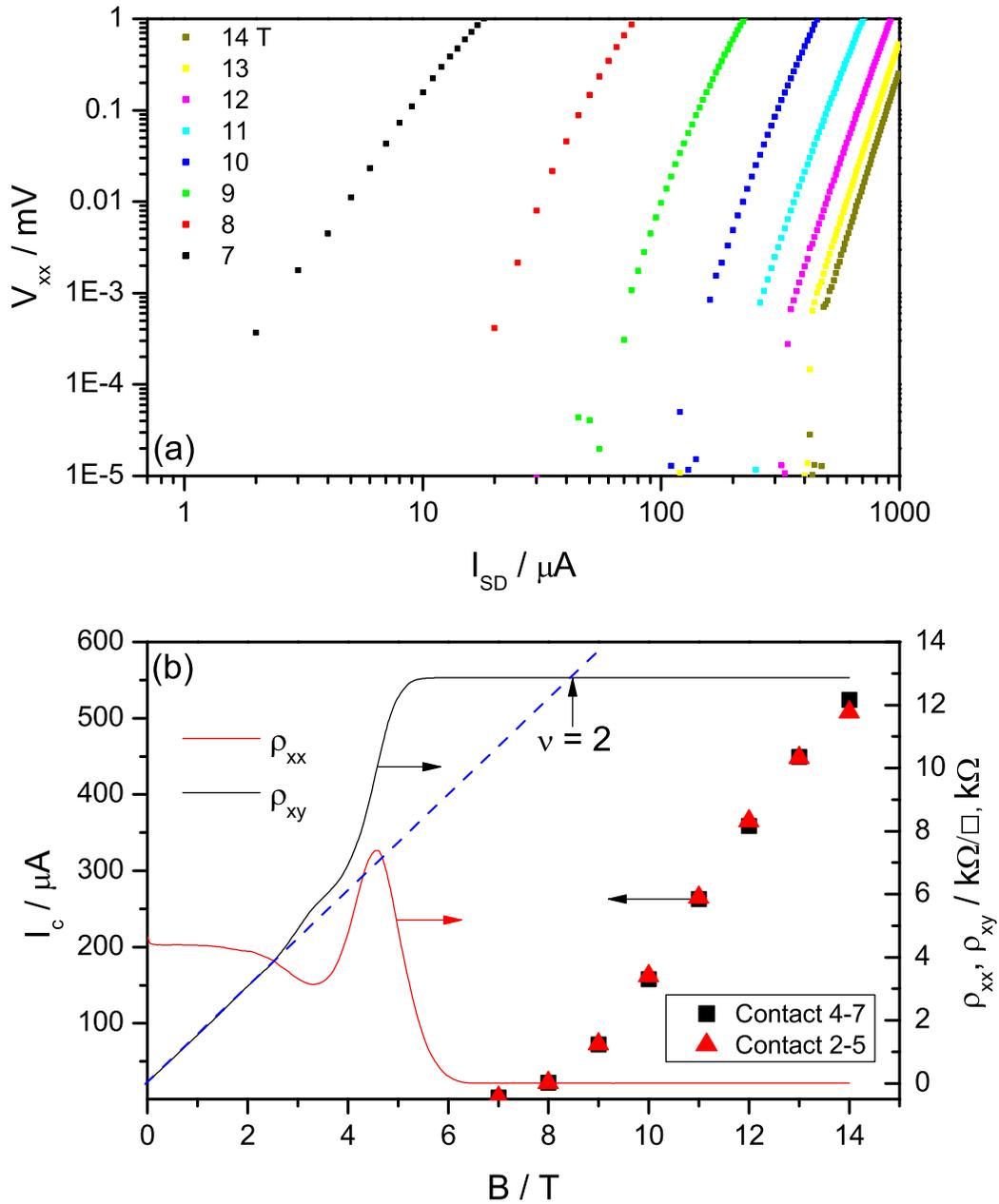}
\caption{\label{fig3} (a) Measurement of $V_{xx}$ (contacts 4 and 7) as a function of source-drain current at different values of magnetic flux density ranging from $B=7\ \rm T$ to (left hand curve) to 14~T (right curve) in steps of 1~T. (b) Transverse ($\rho_{xy}$) (contacts 4 and 5) and longitudinal ($\rho_{xx}$) (contacts 4 and 7) resistivity measurement at the reduced carrier density of $\rm 4.6\times 10^{11}\ cm^{-2}$ measured at $I_{sd}=1\ \rm\mu A$ together with the measured breakdown current, $I_c$. Dashed blue line indicates position of the exact $\nu=2$ filling factor for the low field carrier density.}
\end{figure}

In order to extend the range of applicability the $\nu=2$ quantum Hall plateau can be brought down in magnetic field by reducing the carrier density using a novel photochemical gating technique~\cite{Lara-Avila2011}. This is achieved by coating the device with a polymer bilayer, a spacer layer (PMMA) followed by an active polymer (ZEP520) able to generate potent electron acceptors under UV light exposure. The effect can be reversed by heating the sample to 170$^\circ$C just above the glass transition temperature of the polymer. Using this technique the carrier density of our device was reduced to $\rm 4.6\times 10^{11}\ cm^{-2}$ resulting in an extremely wide $\nu=2$ quantum Hall plateau as can be seen in figure~\ref{fig3}(b).

Accurate quantum Hall resistance measurements require that the longitudinal voltage remains zero (in practice, below the noise level of the voltmeter) to ensure the device is in the non-dissipative state, which can be violated by the breakdown of the QHE at high source-drain current levels. Fig.~\ref{fig3}(a) shows the determination of the breakdown current $I_{c}$ for different values of $B$ along the $\nu=2$ plateau. Here we define $I_{c}$ as the source-drain current, $I_{sd}$, at which $V_{xx}\geq 100\ \rm nV$. Above the breakdown current the voltage increases rapidly as a function of current as a consequence of the sudden increase in electron temperature and subsequent increase in longitudinal resistivity~\cite{Yoshioka2002}. In figure~\ref{fig3}(b) the values of the breakdown current are plotted as a function of $B$ together with $\rho_{xx}$ and $\rho_{xy}$.

\begin{figure}
\includegraphics{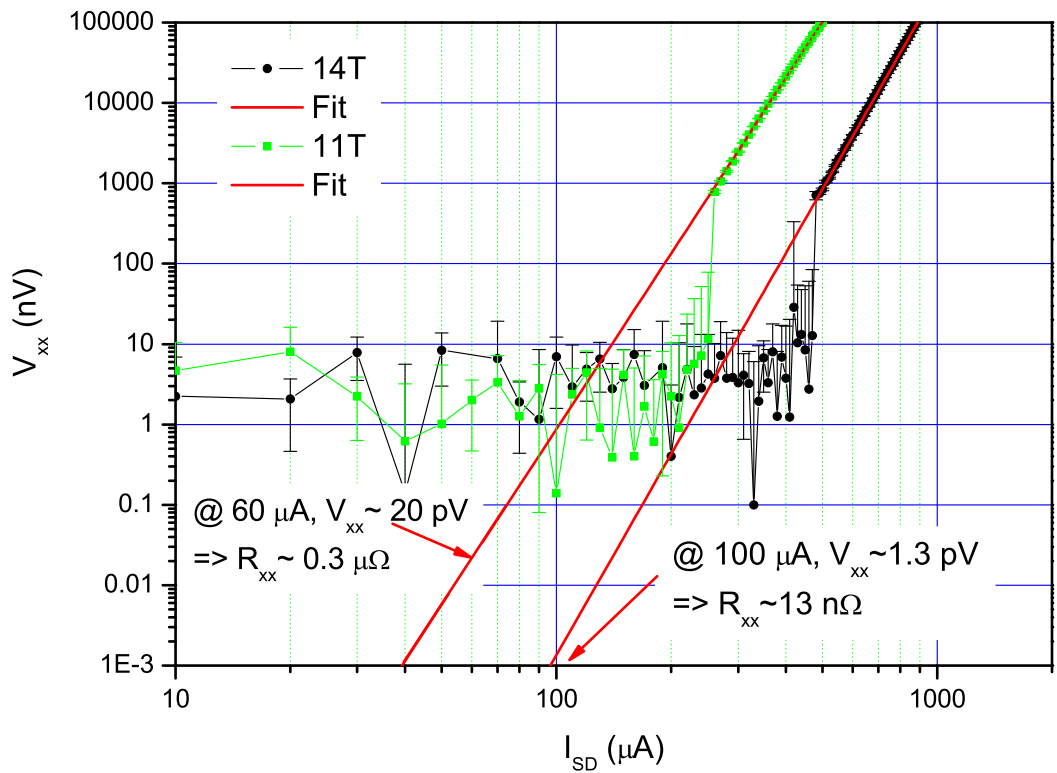}
\caption{\label{fig4} (a) Log-log plot of $V_{xx}$ as a function of source-drain current at 11~T (Green squares) and 14~T (Black dots). The red lines are fits to the data at high source-drain currents. 5~nV is the noise floor of the voltmeter.}
\end{figure}

Figure \ref{fig4} shows the measurement of $V_{xx}$ as a function of $I_{sd}$ at 11 and 14~T in more detail (these are the magnetic flux densities at which the comparison measurements are performed). On the log-log plot it is clear that above the breakdown current the longitudinal voltage shows a power-law dependence on current. The red lines in the figure are a fit to this part of the trace.  For the highest flux density of $14\ \rm T$ the breakdown current is $\sim 500\ \rm\mu A$, however, the maximum source-drain current used in our comparison measurements is $100\ \rm\mu A$. The extrapolated fitted line indicates that the expected $V_{xx}$ is $\sim 1.3\ \rm pV$ at $100\ \rm\mu A$ and therefore $R_{xx}\approx 13\ \rm n\Omega$. Note that close to the breakdown current $V_{xx}$ drops away much more quickly than this exponential behaviour before it disappears in the noise of the DVM ($\pm 5\ \rm nV$) (the shape of the breakdown curve is remarkably similar to that observed by Cage et al.~\cite{Cage1983} for GaAs heterostructure devices) and therefore the $R_{xx}\approx 13\ \rm n\Omega$ is the upper bound. On the basis of this analysis we can be confident that the longitudinal resistivity is vanishingly small and that there is a considerable safety margin on $I_{sd}$ in our experiments. A few measurements were performed at a lower magnetic flux density around 11~T. For these measurements a lower $I_{sd}$ of $60\ \rm\mu A$ was chosen and a smilar extrapolation of the high source-drain currents gives $V_{xx}\approx 20\ \rm pV$ and $R_{xx}\approx 0.3\ \rm\mu\Omega$.

Returning to figure~\ref{fig3}(b), we note that the measured $I_c$ for contact pairs on either side of the device is virtualy identical. $I_c$ as a function of magnetic flux density shows a rather unusual dependence as the breakdown current continues to increase in the $\nu=2$ quantum Hall state reaching $\rm \approx 500\ \mu A$ at our maximum field of $B=14\ \rm T$. This behaviour is very different from that observed in conventional semiconductor systems where the breakdown current peaks at the exact integer filling factor~\cite{Jeckelmann2001} indicated by the dashed blue line in Fig.~\ref{fig3}(b). This behaviour can be explained by a magnetic field dependent charge transfer mechanism between the interfacial layer and graphene~\cite{Janssen2011a}. The amount of charge transfer is determined by the interplay of classical (as a result of the geometry) and quantum (as a result of the density of states) capacitance of the device structure. The density of states oscillates as the Landau levels (LL) pass through the Fermi level when the magnetic field increases. The result is that the carrier density in the graphene layer is also oscillating and specifically when the Fermi level is between the $N=0$ and $N=1$ LL, the carrier density increases linearly with magnetic field. This effectively pins the filling factor at $\nu=2$ over a broad range of magnetic field resulting in {\it a novel quantum Hall state which is ideally suited for high precision resistance metrology}. The anomalous pinning is responsible for the {\it extremely high breakdown current} and {\it wide operational parameter space} of an epitaxial graphene quantum Hall device.

\begin{figure}
\includegraphics{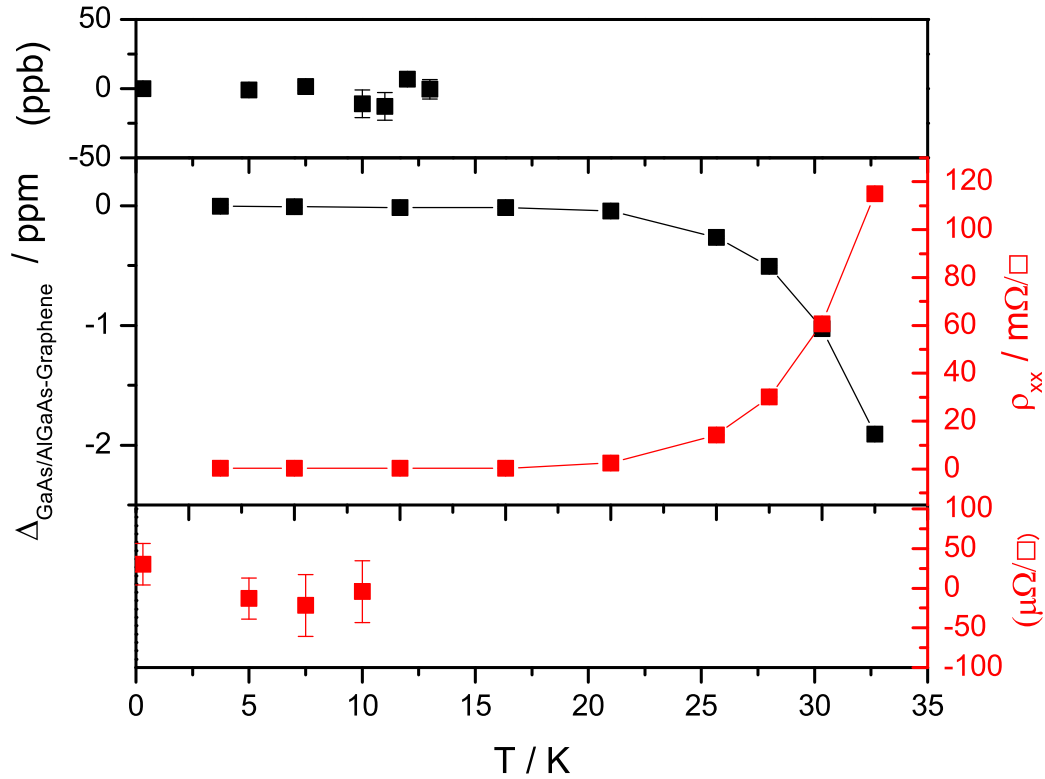}
\caption{\label{fig5} Middle: Measurement of 
$\Delta_{\rm GaAs/AlGaAs-Graphene}=[R_H({\rm GaAs/AlGaAs}, T=1.5\ {\rm K})-R_H({\rm Graphene}, T)]/(R_{\rm K}/2)$ 
and $\rho_{xx}$ as a function of temperature for the graphene device. $B=14$~T for the graphene device and 10.5~T for the GaAs/AlGaAs device. The measurement current was $\rm 60\ \mu A$. Top and bottom: high-resolution measurements of $\Delta_{\rm GaAs/AlGaAs-Graphene}$ and $\rho_{xx}$ demonstrating ppb-level quantisation. For the middle panel the carrier density was $4.6\times 10^{11}\ \rm cm^{-2}$ and for the top and bottom panel the density was $6.7\times 10^{11}\ \rm cm^{-2}$.}
\end{figure}

Figure~\ref{fig5} demonstrates the robustness of the $\nu=2$ quantum Hall state as a function of temperature. Here 
$\Delta_{\rm GaAs/AlGaAs-Graphene}=[R_H({\rm GaAs/AlGaAs}, T=1.5\ {\rm K})-R_H({\rm Graphene}, T)]/(R_{\rm K}/2)$ quantifies the difference between the graphene sample and a fixed reference GaAs/AlGaAs device (the measurement technique is explained in detail in section~\ref{ccc}). In a separate measurement the longitudinal resistivity was measured for the graphene device and plotted in the same graph. The temperature for the reference device is held constant at 1.5~K while the temperature for the graphene device is varied. The middle panel of fig.~\ref{fig5} shows a low resolution measurement across a wide temperature range and demonstrates that $R_H$ starts to deviate as soon as a measurable $\rho_{xx}$ appears. The top and bottom panels are high resolution measurements which demonstrate ppb-level quantisation up to 15~K when $\rho_{xx}$ is of the order of several tens of $\rm \mu V$'s (the high resolution measurements where performed on an earlier cooldown cycle of the device when the charge carrier density was higher).  

\begin{figure}
\includegraphics{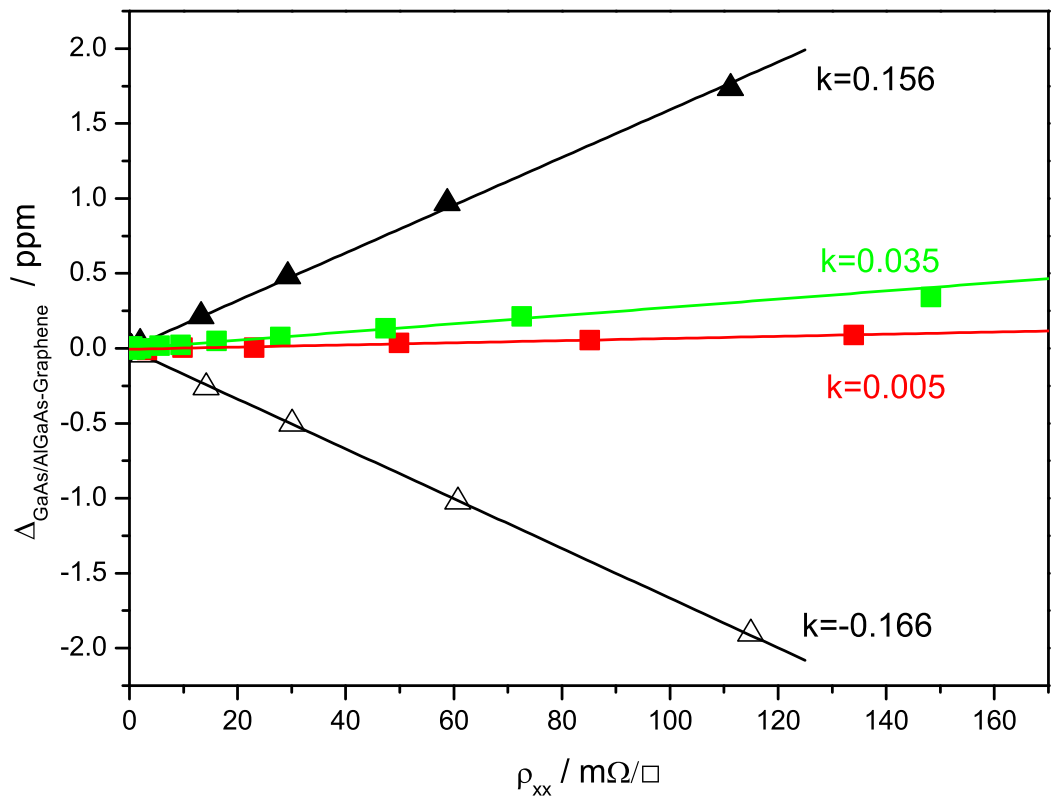}
\caption{\label{fig6} Variation of $\Delta_{\rm GaAs/AlGaAs-Graphene}$ as function of $\rho_{xx}$ for three different charge carrier densities determined from data such as displayed in fig.~\ref{fig5}. Red = $9.1\times 10^{11}\ \rm cm^{-2}$, Green = $6.7\times 10^{11}\ \rm cm^{-2}$ and Black = $4.6\times 10^{11}\ \rm cm^{-2}$. Open black triangles are measured in reverse field direction. Solid lines are fits to the data.}
\end{figure}

In figure~\ref{fig6} we have plotted $\Delta_{\rm GaAs/AlGaAs-Graphene}$ versus $\rho_{xx}$ for three separate cooldown cycles of the device. Each time $\Delta_{\rm GaAs/AlGaAs-Graphene}$ is proportional to $\rho_{xx}$, identical to the well-known empirical relationship $\Delta R_{H}=k\rho_{xx}$ which has been observed for traditional semiconductor systems~\cite{Jeckelmann2001}. The value of $k$ varies significantly between different runs which are signified by different charge carrier densities (in between cooldowns the sample was exposed to UV radiation in order to reduce the carrier density in small controllable steps). For the final and lowest charge carrier density we also measured $k$ in opposite magnetic field direction. The fact that the sign of $k$ changes with the $B$-field direction indicates that the observed relation between $R_H$ and $\rho_{xx}$ is due to inhomogeneity of the charge density in the sample as proposed by van der Wel {\it et al.}~\cite{vanderwel1988} rather than the finite width of the voltage probes for which no sign change is expected. In the latter case $k=\frac{l_p}{W}\approx 0.7$ for both $B$-field directions which is clearly inconsistent with the experimental data. A plausible explanation for the increase in $k$ with reducing carrier density could be the increase in inhomogeneity in the sample as we get closer to the Dirac point~\cite{Novoselov2005}. This is supported by the fact that the value of $k$ was found to be different for different combinations of measurement contacts (not shown). 

Combining the results from figure~\ref{fig4} where we estimated $R_{xx}$ with the value of $k$ obtained in figure~\ref{fig6} allows us to estimate the expected relative error in $R_H$. For $B=14\ \rm T$ and $I_{sd}=100\ \rm\mu A$ we obtained $R_{xx}=13\ \rm n\Omega$ which with $k=0.16$ implies that we can expect a relative error in $R_H$ much less than $10^{-12}$. Similarly for $60\ \rm\mu A$ at $B=11\ \rm T$ the relative error would be much less than $10^{-11}$. 

\begin{figure}
\includegraphics{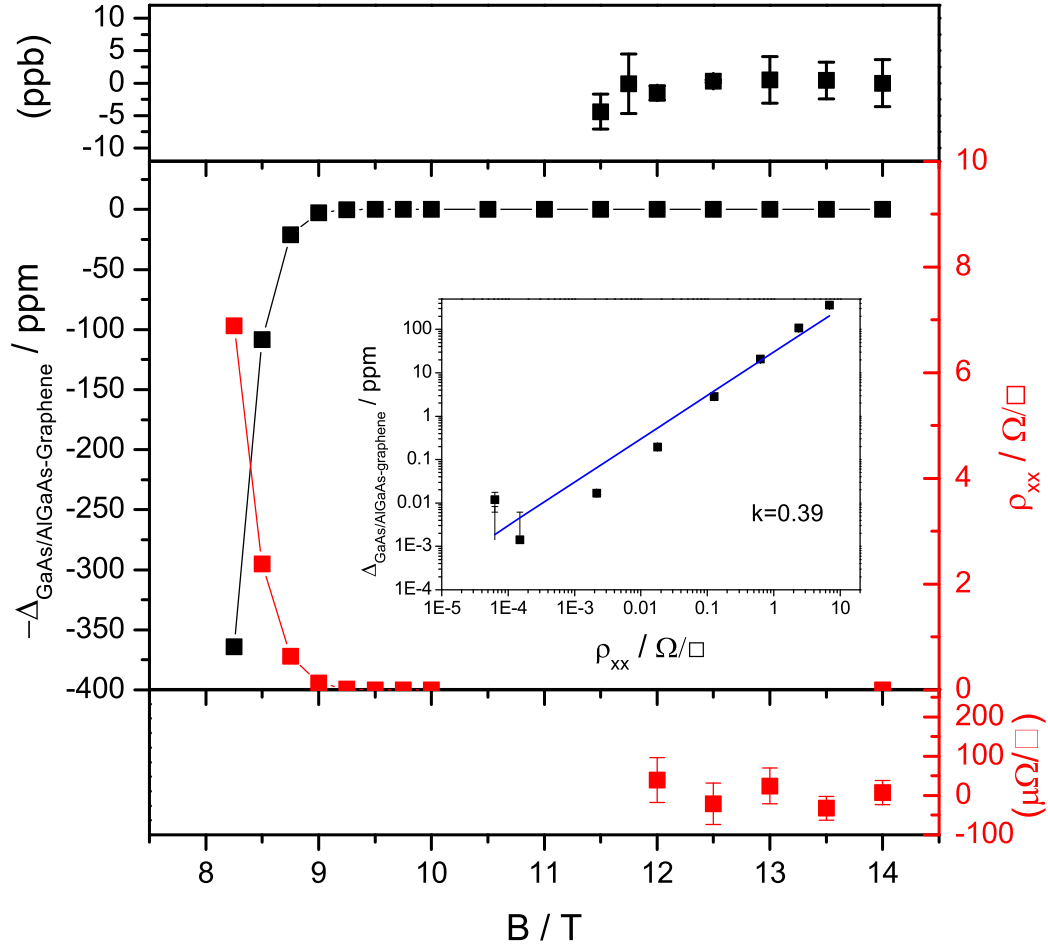}
\caption{\label{fig7} Middle: Measurement of $-\Delta_{\rm GaAs/AlGaAs-Graphene}=[R_H({\rm GaAs/AlGaAs}, B=10.5\ {\rm T})-R_H({\rm Graphene}, B)]/(R_{\rm K}/2)$ and $\rho_{xx}$ as a function of $B$ on the graphene device. $T=0.3\ \rm K$ for the graphene device and 1.5~K for the GaAs/AlGaAs device. The measurement current was $60\ \rm\mu A$. Top and bottom: high-resolution measurements of $-\Delta_{\rm GaAs/AlGaAs-Graphene}$ and $\rho_{xx}$ demonstrating ppb-level quantisation. For the middle panel the carrier density was $4.6\times 10^{11}\ \rm cm^{-2}$ and for the top and bottom panel the density was $6.7\times 10^{11}\ \rm cm^{-2}$. Inset: log-log plot of the variation of $\Delta_{\rm GaAs/AlGaAs-Graphene}$ as function of $\rho_{xx}$ for the data in the middle panel.}
\end{figure}

In figure~\ref{fig7} (middle) the same experiment as in figure~\ref{fig5} is repeated but this time with magnetic flux density as the parameter and $T=300\ \rm mK$. The magnetic flux density for the reference device is held constant at the centre of the $\nu=2$ plateau and the magnetic flux density for the graphene sample is varied. Top and bottom panel are high resolution measurements which demonstrate ppb-level quantisation (again measured on a different cooldown cycle). The graph shows that the $\nu=2$ quantum Hall plateau extends over a range of at least 4~T and the measurement is only limited from above by the maximum available magnetic field. The inset is a log-log plot of $\Delta_{\rm GaAs/AlGaAs-Graphene}$ as a function of $\rho_{xx}$ similar to fig.~\ref{fig6} and the blue line is $\Delta R_H=k\rho_{xx}$ with $k=0.39$. From the plot it can be seen that the data do not quite follow a linear dependence as might be expected. A likely reason for this deviation is the fact that as a function of $B$ the charge carrier density in epitaxial graphene is not constant which could result in a variation of $k$ as a function of $B$. 

Comparing the performance of our graphene device with that of GaAs devices shows a significantly wider operational parameter space for the $\nu=2$ quantum Hall state. In GaAs the field range is usually a few tenths of teslas up to 1 Tesla for the best devices and for temperature the operational range is generally below 2~K~\cite{Jeckelmann2001,Poirier2009}. This result is not surprising given the fact that the energy spacing between the $n=0$ and $n=1$ Landau levels is approximately seven times larger in graphene than in GaAs around a magnetic flux density of 10~T. Also the breakdown current in graphene compares favorably with that in GaAs. Although a breakdown current as large as $500\ \rm\mu A$ has been obtained in mesoscopic GaAs devices, our device is only $35\ \rm\mu m$ wide, at least a factor of ten narrower than standard Hall bars used for metrological measurements. This leaves significant scope for further improvement if wider, homogeneous, graphene devices can be produced. It is important to note that for the lowest charge carrier density ($\sim 4\times 10^{-11}\ \rm cm^{-2}$) in our graphene device we have not yet reached the centre of the $\nu=2$ quantum Hall plateau and the operational parameter space is likely to be even wider at higher magnetic flux densities or in devices with even lower charge carrier density. 

\section{Characterisation of GaAs samples}\label{CharGaAs}
The GaAs/AlGaAs heterostructures used in this work were standard Hall bar devices in use for routine quantum Hall metrology for the last 20 years at NPL and BIPM. One device was supplied originally by PTB (marked as PTB2) and the other by Laboratoires d'\'Electronique Philips (LEP). The dimensions and characteristic parameters are listed in Tab.~\ref{tab2} together with those of the graphene device discussed in the previous section. 

\begin{table}
	\caption{Device parameters. The dimensions are defined in Fig.~\ref{fig1}. The contacts on the PTB2 sample were made of small tin balls at the edges of the chip and so voltage probe width is not applicable.}
	\label{tab2}
	\centering
		\begin{tabular}{lllllll}
		\br
			Device & $L$    & $W$    & $l$    & $l_p$ & $\mu$          & $n_s$                        \\
						 & (mm)   & (mm)   & (mm)   & (mm)  & ($\rm T^{-1}$) & ($\rm \times 10^{11}\ cm^{-2}$) \\
			\mr
			PTB2 	 & 6.0    & 2.5    & $\sim 1.5$      & N/A     & 40             & 4.6                        \\
			LEP 	 & 2.2    & 0.4    & 0.5      & 0.050       & 50             & 5.1                          \\
			GR9 	 & 0.16   & 0.035  & 0.06   & 0.024   & 0.75           & 4.6                          \\
			\br
		\end{tabular}
\end{table}

Before commencing the high-accuracy measurements both GaAs/AlGaAs samples were characterised according to the guidelines for quantum Hall resistance metrology~\cite{Delahaye2003} (i.e. we confirmed that the three-terminal contact resistance measured on the $\nu=2$ plateau was of the order of a few ohms for all contacts used and that the longitudinal resistivity at the measurement current was below $\rm 10\ \mu\Omega$). The breakdown current, $I_c$, measured at 1.5~K was  $\rm 150 \ \mu A$ for the PTB2 device and $\rm 100 \ \mu A$ for the LEP device.

\section{Cryogenic current comparator bridge}\label{ccc}

\subsection{Design principle}

\begin{figure}
\includegraphics{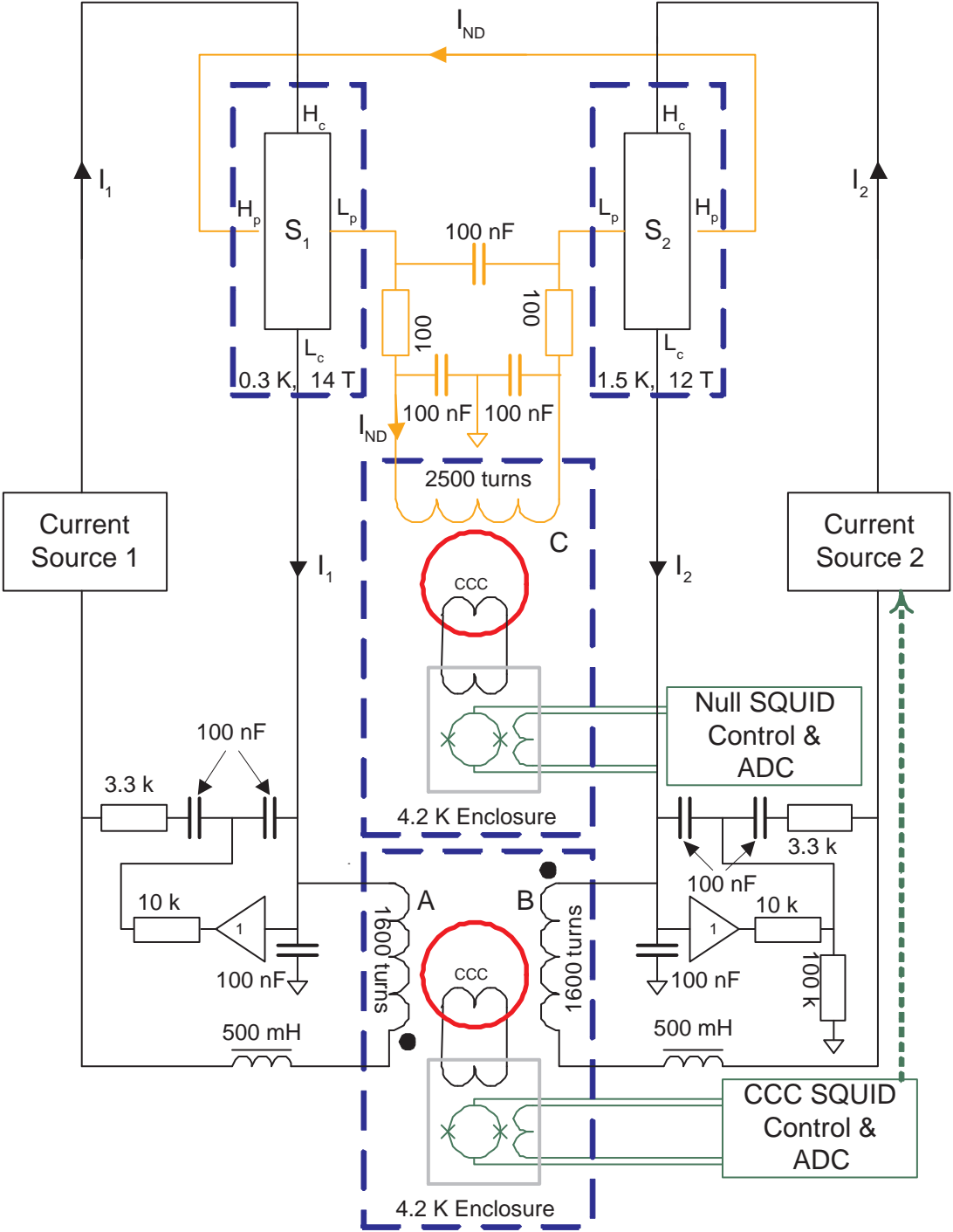}
\caption{\label{fig8} Schematic of the cryogenic current comparator bridge circuit}
\end{figure}

The measurements were made with a cryogenic current comparator (CCC) bridge which has been described in detail in Refs.~\cite{Williams2010,Williams2011} and is illustrated in simplified form in Fig.~\ref{fig8}. Isolated current sources 1 and 2 separately drive current through samples $S_1$ and $S_2$ and associated windings A and B on the CCC. The current ratio can be set via electronics to a few parts in $10^6$ and this ratio is improved to a level of 1 part in $10^{11}$ by forming a negative feedback loop from the SQUID (Superconducting Quantum Interference Device) sensing the net flux in the CCC to one of the current sources. Our standard CCC uses the model A20 nanovolt amplifier manufactured by EM Electronics. The performance of this amplifier can be characterised as a voltage noise with Allan deviation of 0.14~nV and a current noise with Allan deviation of 0.2~pA for a 10~s observation (also see Fig.~4 in Ref.~\cite{Williams2011}). When measuring two quantum Hall devices with a combined impedance of $\approx 25\ \rm k\Omega$ this current noise corresponds to 5~nV and  dominates the resolution of the measurement system. We therefore chose to use a second CCC as a null detector \cite{Delahaye1991}. The potential contacts on $S_1$ and $S_2$ are closed in a loop via winding C on this second CCC. This device is configured with just a single winding to measure a current null rather than two windings to establish a current ratio. 

The CCC for the bridge current ratio is fitted with an RLC filter which is a modified version of the RC filter described in~\cite{Williams2010}. The modification gives a lower bandwidth of 1~kHz and was found to be necessary due to the larger interference experienced in this experiment involving four cryostats and long connecting cables between laboratories.  Also, as a much higher accuracy was required for this precision comparison of quantum Hall samples than is needed for routine resistance measurement a guard, driven by a unity-gain operational amplifier, was added to the capacitors in the filter on each side of the bridge~\cite{Williams1991} to reduce any leakage across the CCC windings to a negligible level. The CCC for null detection was also fitted with a filter (at room temperature) to limit the bandwidth to approximately 16~kHz. The filter was constructed using $100\ \Omega$ resistors so as to keep the additional Johnson noise in the null detector circuit at the nV level. The potential difference between the bridge circuit and the overall measurement system screen in the region of the null detector is held close to zero by the 100~k$\Omega$ resistor connected to the screen in the filter for current source 2. Any leakage resistance in the filter capacitors will therefore have a negligible effect and a guard circuit for these is not necessary.

The parameters for both CCC's are listed in Tab.~\ref{tab3}. The current sources and SQUID control units are connected to a computer which supervises the setting of the measurement currents and collects data from the two SQUIDs. Data are collected alternately in forward and reverse current direction so as to eliminate electrical offsets and drift. 

\subsection{Bridge sensitivity}

The CCC bridge sensitivity can be derived from the schematic network in Fig.~\ref{fig8}. At bridge balance the flux in the ratio CCC is zero so that $I_1 = I_2 = I$. A deviation between the ratio $R_1 / R_2$ from unity will result in a small  current, $I_{ND}$ in the null detector circuit. We have,

\begin{eqnarray}
R_1(I-I_{ND})=R_2(I+I_{ND}),\label{eq1}\\
\frac{R_2}{R_1}=\frac{I-I_{ND}}{I+I_{ND}}\approx1-\frac{2I_{ND}}{I}.\label{eq2}
\end{eqnarray}

The sensitivity of the null detector was checked by placing a $10\ \rm G\Omega$ shunt resistor in parallel with one of the quantum Hall devices. For a measurement current of $100\ \rm\mu A$ a SQUID signal for the null detector CCC corresponding to $I_{ND} = 70\ \rm{pA}$ was observed which, according to equation~\ref{eq2}, corresponds to a deviation of the resistance ratio from unity of $1.4 \times 10^{-6}$. This is in reasonable agreement with the calculated deviation of $1.3 \times 10^{-6}$, confirming the null detector sensitivity.  A second feedback loop can be operated to bring the SQUID signal from the null detector to zero using a third winding on the CCC~\cite{Williams2010}. However, this is only necessary when an accurate resistance ratio measurement for resistors with deviations from nominal is required. In this experiment we are comparing two quantum Hall devices which are expected to be exactly equal to a very high degree, so this second feedback loop is not necessary and the open loop output of the null detector is simply recorded instead.  

\begin{table}
	\caption{Summary of individual noise components in the bridge circuit and total expected noise. The noise attributed to the two CCCs is measured at 10~s and converted to voltage noise across one QHR device in case of the current CCC and across two QHR devices for the voltage CCC. The noise arising from the resistors is calculated. Last two columns are the Allan Deviation (AD) at a 10 second measurement time. Note that for white noise of power spectral density $h_0\ \rm V^2Hz^{-1}$ the corresponding Allan deviation is $(h_0/2\tau)^{-1/2}$.}
	\label{tab3}
	\centering
		\begin{tabular}{llllll}
		\br
			Component 	& Comment    				& Sensitivity   			   			      & CCC null      & A20 null     \\
			            &                   &                                   & $\rm AD$ (nV) & $\rm AD$ (nV)\\
			\mr
			Ratio CCC 	& 1600:1600 t  	  	& $\rm 16\ \mu A\cdot$turn$\phi_0^{-1}$   			  & 1.3           & 1.3   			  \\
			Null CCC  	& 2500 t   					& $\rm 13\ \mu A\cdot$turn$\phi_0^{-1}$     			& 1.3           &             \\
			A20 null    & Voltage noise     &                                   &               & 0.14        \\
			A20 null    & Current noise     &                                   &               & 5.0        \\
			QHR1 	 			& 300 mK   					& 			                       			& 0.1           & 0.1           \\
			QHR2 	 			& 1.2 K   					& 					                      	& 0.2           & 0.2            \\
			Filter 	 		& $200\ \Omega$  		& 					                      	& 0.4           &              \\
			\br
			Total				&										&											  						& 1.9	          & 5.2          \\
			\br
		\end{tabular}
\end{table}

\subsection{Noise and measurement resolution}
The ultimate resolution of the bridge can be estimated by considering the noise components due to the current balance CCC, the 
null detector CCC and filter network, and the Johnson noise from the quantum Hall devices being measured. Here we use the Allan deviation of the time domain signals~\cite{Witt2000,Williams2010} to analyse the various components. The design of the bridge is based on a noise performance of the DC SQUID and CCC combination with an Allan deviation of $10\ \mu\phi_0$ for a 10~s measurement interval. Table~\ref{tab3} lists the parameters for the different components of the bridge and the calculated voltage noise in the null detector loop in terms of the Allan deviation for a 10~s measurement interval. The total noise voltage for the bridge with CCC nanovoltmeter is 1.9~nV compared to 5.2~nV when using the A20 nanovoltmeter. 

\begin{figure}
\includegraphics{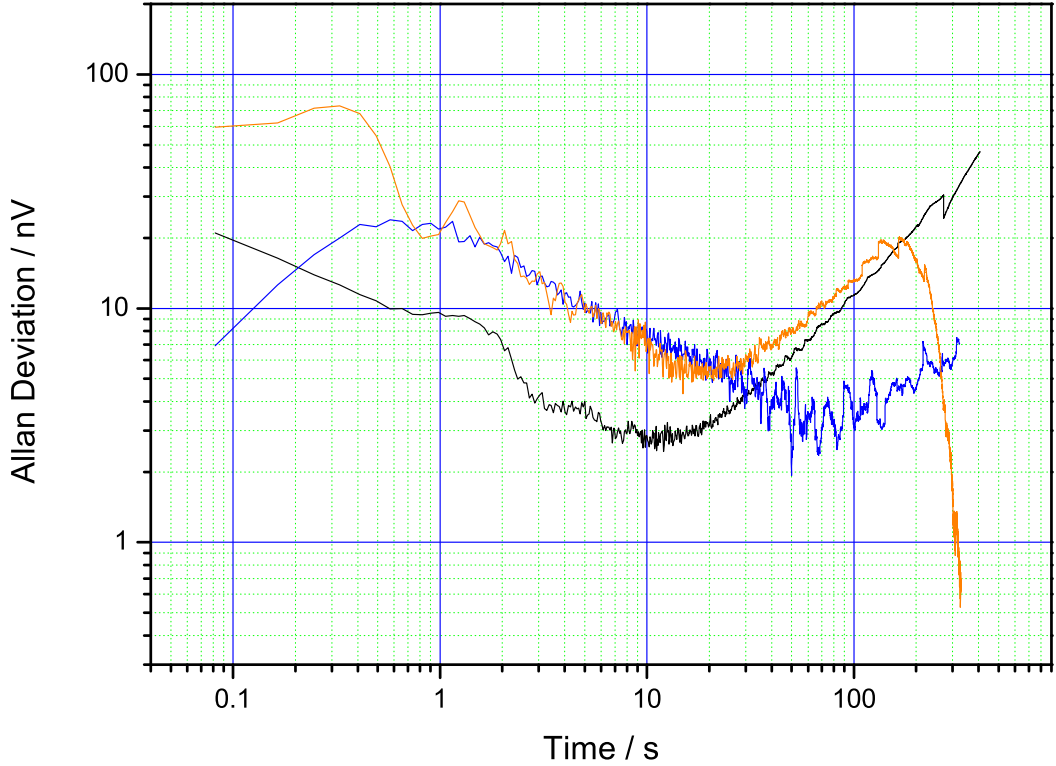}
\caption{\label{fig9} Allan deviation of the CCC nanovolt meter (orange) and A20 nanovoltmeter (blue) when two QHR devices are connected to the bridge. Also shown is SQUID noise of the current balance CCC which was scaled to 1600 turns.}
\end{figure}

Figure~\ref{fig9} shows the measured Allan deviation of the CCC null detector with two QHR devices connected to the bridge (the master and slave current sources and current balance CCC are disconnected). It is again expressed as a voltage noise in the null detector circuit. It can be seen from this graph that at 10~s the nanovoltmeter noise is approximately 8~nV compared to design value of 1.3~nV. For comparison the measured Allan deviation is also shown for the A20 null detector which is similarly $\sim$8~nV at 10~s rather than the expected 5~nV. This suggests that both the CCC null detector and the A20 instrument are experiencing a similar level of excess noise. The enhancement of the noise is not surprising given the fact that we are using two quantum Hall systems each with their own magnet power supply and peripheral pumping equipment. In addition the connecting cables between these systems and the measurement bridge span a distance of more than 10 meters across two laboratories. 

In Fig.~\ref{fig9} the noise from the current balance CCC is also shown and has been scaled for a 1600 turn winding and a single quantum Hall resistance to give an equivalent voltage noise in the null detector circuit. The Allan deviation goes through a minimum of $\sim$~3nV at about 10~s which is again more than the expected 1.3~nV and then increases at one decade per decade in sampling interval. (Note that the minimum in the Allan deviation determines the optimum data acquisition time of $\sim 10\ \rm s$ in our measurement system.) Both SQUIDs exhibit a similar behaviour for time intervals longer than 30~s. The excess noise in the null detector circuit is clearly the limiting factor in our measurement system at present. Excess noise has also been seen in measurement systems of this type \cite{Delahaye1991} and whereas in a manual data acquisition system the user can select apparently lower noise results, here our system runs for many hours unattended with the consequence that all data points are included in our analysis. From Fig.~\ref{fig9} we can conclude that the optimal measurement time is approximately 20~s for a given current direction. For longer times both the CCC null detector noise and current balance CCC noise contributions will start to increase with the null detector at present dominating until time intervals longer than 50~s. A future improvement in this experiment would be to select a graphene and GaAs device with coinciding $\nu=2$ plateaux which can be co-located in the same cryostat and thereby reduce the complexity of the connecting cables.

\begin{figure}
\includegraphics{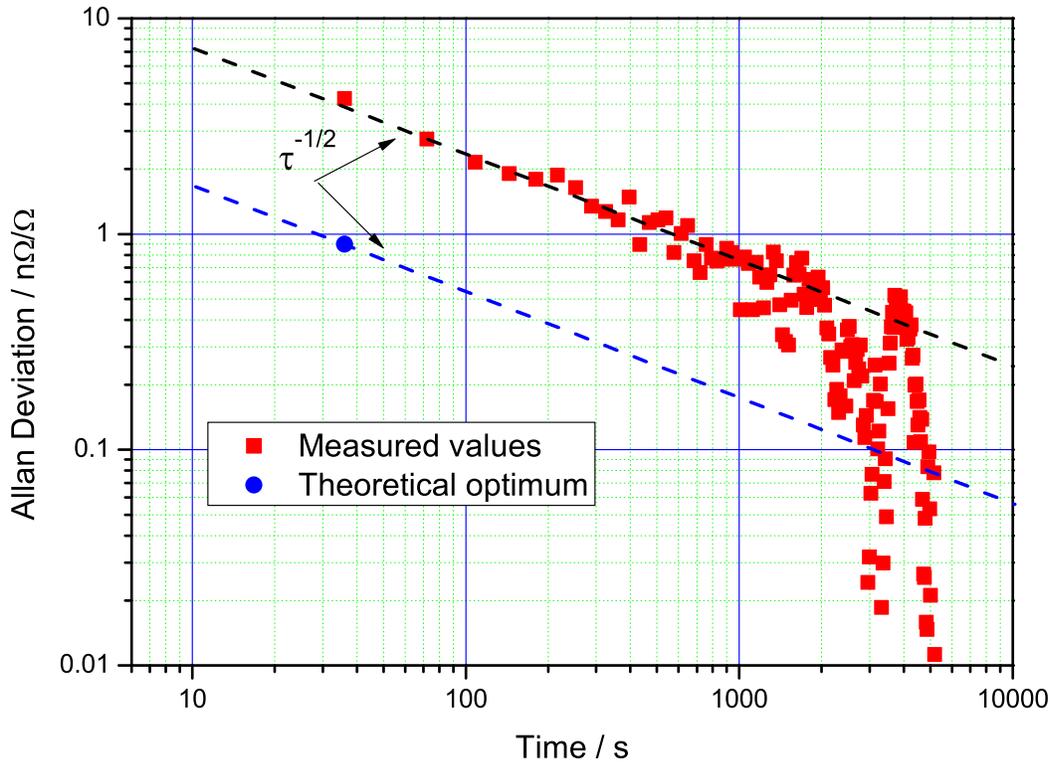}
\caption{\label{fig10} Typical Allan deviation of the resistance ratio for a 1:1 measurement of graphene against GaAs/AlGaAs heterostructure at $\nu=2$ for a measurement current of $100\ \rm\mu A$.}
\end{figure}

Figure~\ref{fig10} shows the Allan deviation for a 3.5~h measurement of graphene against GaAs at $\nu=2$. The measurement data have been analysed in blocks of three intervals consisting of a 1~s settling time plus 10~s data collection interval (for optimum CCC performance) and a 1~s data transfer each to give a value for the resistance ratio every 36~s. (Note that the settling time for the CCC nanovoltmeter is much faster than that of the A20 where we have to allow for at least 10~s settling time \cite{Williams2010}. We regularly checked the stability of the servo loop by varying the wait-time after current reversal and inspecting the measurement result for systematic correlations.) Adding all the noise components in Tab.~\ref{tab3} in quadrature and using the observed noise for the CCC nanovoltmeter (which dominates) and current balance CCC in Fig.~\ref{fig9} gives a total noise of 8~nV for a 10~s measurement interval. For a measurement current of $\pm 100\ \rm\mu A$ in $R_{\rm K, \nu=2}$ this corresponds to a relative deviation of 3~parts in $10^9$. This number can be related to a block of three intervals by multiplying by a scaling factor of $\sqrt{3/2}$ \cite{Williams2010} to give 4~parts in $10^9$ for the expected Allan deviation. From Fig.~\ref{fig10} we see that the first data point is also at 4~parts in $10^9$ confirming that the overall measurement resolution is consistent with the measured noise. The Allan deviation decreases as $1/\sqrt{\tau}$ as expected for white noise and shows that the current reversal technique efficiently removes the non-white noise visible in Fig.~\ref{fig9}. After 3.5~h measurement time a relative uncertainty of 2 parts in $10^{10}$ is achieved. If the two CCC's can be made to operate at their optimum noise performance, a single block of three measurements would give a relative uncertainty slightly better than 1 part in $10^{9}$ and 6 parts in $10^{11}$ after 3.5~h (blue dot and dashed line in Fig.~\ref{fig10}). 

\section{Uncertainty analysis}

\subsection{Type B analysis}\label{TypeB}

\subsubsection{Ratio error}
The ratio error is determined by connecting both 1600 turn windings in series opposition. A large (10~mA) current is passed through the windings and the SQUID signal recorded (total applied current linkage is $32\ \rm A\cdot$turn). This process is repeated several hundred times in both forward and reverse current direction to improve the measurement resolution. A least squares fit to the data gives $17\pm 7\ \rm\mu V$ for the residual SQUID signal. The sensitivities of the SQUID and CCC are $0.7\ {\rm V\phi_0^{-1}}$ and $16\ \rm\mu A\cdot$turn$\phi_0^{-1}$, respectively, giving an imbalance of $3.6\pm 0.9\times 10^{-10}\ \rm A\cdot$turn. In principle it is possible to correct the measurements results for this imbalance because it is a fixed property of the CCC. However, given the small size we have not done this and have used the imbalance as an uncertainty component, resulting in a relative standard uncertainty of $1.1\times 10^{-11}$ in the uncertainty budget.

\subsubsection{Leakage}
Leakage resistances must be controlled to a very high level in precision resistance measurements depending on where they appear in the bridge circuit. The most stringent condition occurs for the wiring connecting the device in the cryogenic system to the bridge. A leakage between the high and low potential side of the device would appear as a resistance in parallel with the quantum Hall device and result in a direct error. Given that the quantum Hall resistance at $\nu=2$ is approximately $10^4\ \Omega$, this implies a leakage $\geq 10^{15}\ \Omega$ for a relative error less than $10^{-11}$. Measuring such a large resistance on long leads (there is no screen between high and low potential leads for $\sim 3\ \rm m$ from the top of the cryostat probe to the sample in the 14~T/300~mK system at NPL) is not straightforward. Using a Keithley 6430 electrometer and a repeated $\pm 100\ \rm V$ excitation during a 10 hour measurement resulted in $0.1\pm 0.06\ \rm pA$, giving a relative standard uncertainty of $1\times 10^{-11}$ in the uncertainty budget. 

\subsubsection{Servo error}

Any residual error in the current ratio between master and slave will translate directly into a false reading on the null detector. In order to test and correct for this error, we record the residual SQUID voltage from the current balance CCC during measurements. From this we can calculate the resulting current in the null detector CCC via,

\begin{equation}
\label{eq5}
I_{SQ}^{ND}=\alpha V_{SQ}\ \ \alpha=\frac{16\ \rm \mu A\cdot turn\phi_0^{-1}}{2\times 1600\times 0.7\ \rm V\phi_0^{-1}}\approx 7\times10^{-9}\ \rm AV^{-1}. 
\end{equation}

and apply a correction. Typically we find corrections in the range of $\sim3\pm 0.6$~fA which corresponds to a relative correction of $3\pm 0.6$ parts in $10^{11}$ to the resistance ratio for a measurement current of $100\ \rm\mu A$. The reason this small error arises is that the slave current source drifts very slightly relative to the master source after each current reversal, due to self-heating in the current source components. The feedback servo has a single integral term so a linear drift in the current source translates into a steady error voltage at the output of the SQUID. For our system, the residual error current in the CCC of $\sim3$~fA quoted above corresponds to an open loop drift in the current sources of approximately 1~part in $10^6$ over the period of a 20~s measurement.

\subsubsection{Miscellaneous}
In the comparison measurements are performed in sets of ten intervals of successive forward and reverse current (rather than the shorter blocks of three intervals which were used in the previous section to analyse the time evolution of the Allen deviation). Each data set is written to disk and a new measurement started. Typically this operates for many hours and hundreds of data sets are collected. Each data set of ten intervals is analysed using a least squares fitting routine to give a value for the resistance ratio and standard deviation. The residuals of the fit are analysed in order to spot potential problems with the system such as excessive settling time, noise and interference. 

\begin{figure}
\includegraphics{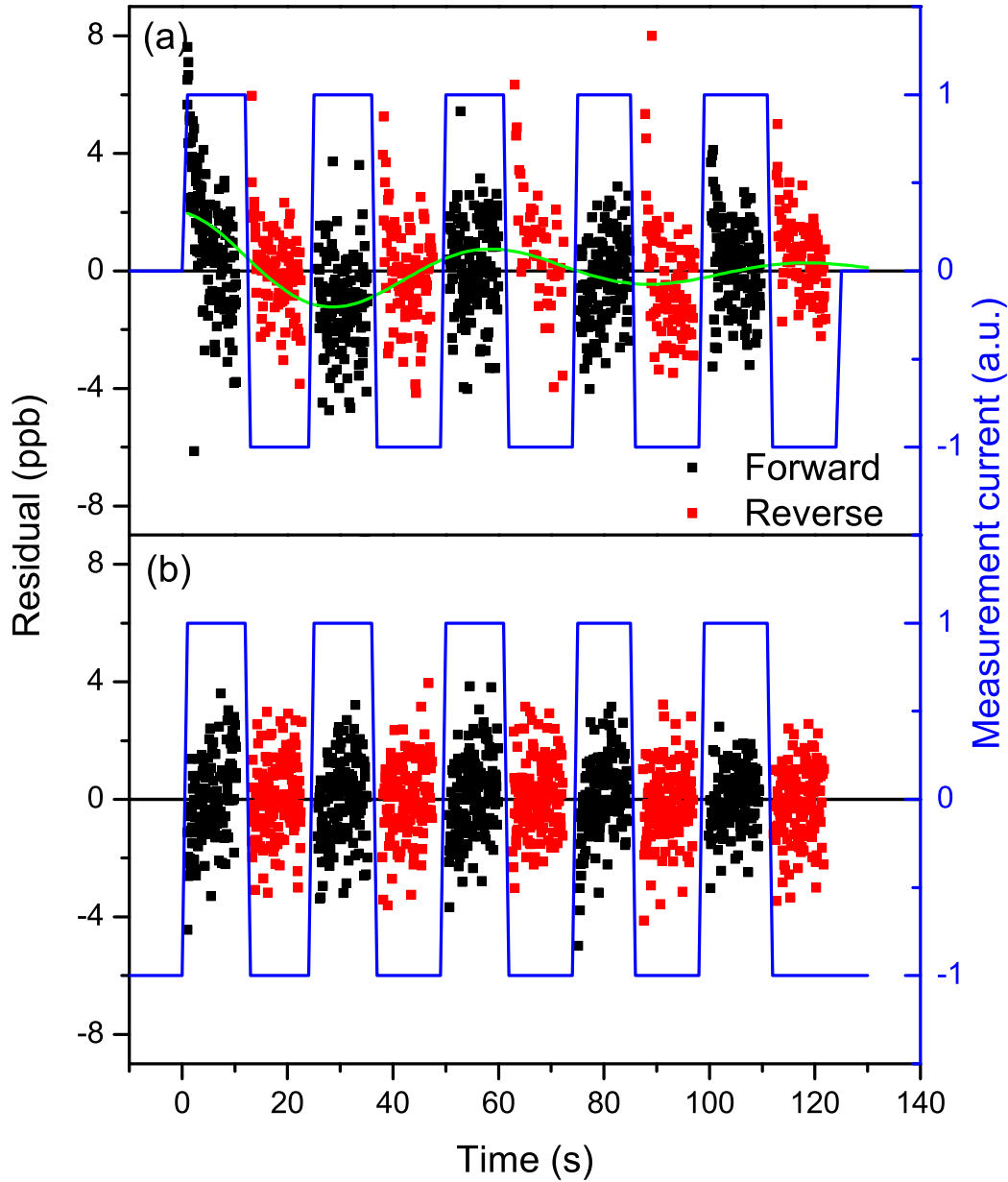}
\caption{\label{fig11} (a) Average of fitting residuals for 300 measurement sets using our standard measurement sequence (blue curve) and (b) modified sequence. Green curve in (a) is guide to the eye.}
\end{figure}

Figure~\ref{fig11}(a) shows the average of 300 sets of fit residuals (i.e. 10 hours total measurement time) and a marked oscillation becomes visible in the data (indicated by the green curve). The oscillation starts at the first forward current measurement interval and damps out with a time constant of about 60~s. We speculate that this effect is caused by a slight heating of the CCC helium dewar due to the measurement current. Normally we start a measurement set from zero current and ramp to maximum forward current, collect data, ramp to maximum reverse current, etc [blue curve in Fig.~\ref{fig11}(a)]. A current reversal takes approximately 1~s. At the end of a measurement set the current is ramped back to zero, data written to file and a new measurement started which combined takes about 6~s, i.e. the current is zero for 6~s between measurement sets. Because the down-leads to the CCC are not superconducting, dissipation occurs during the measurement, and drops for 6~s between measurement sets, causing a temperature-pressure oscillation in the dewar. Averaging the residuals in Fig.~\ref{fig11}(a) results in an error of $\sim 1\times10^{-10}$ which is not insignificant for our comparison.

Figure~\ref{fig11}(b) displays the average of a similar 10 hour measurement where the measurement current was always present in the CCC (i.e. the measurement current was left at maximum reverse current at the end of a measurement set and the new set would start by sweeping from reverse to forward maximum current). As can be seen from this graph the oscillation has disappeared and the mean of all residuals averages to $3\times10^{-15}$ and any remaining error is therefore negligible. 

This effect demonstrates the care that must be taken when making very high precision measurements and the power of automated measurements and data analysis in tracking down extremely small systematic errors. 

\section{Type A analysis}
\begin{figure}
\includegraphics{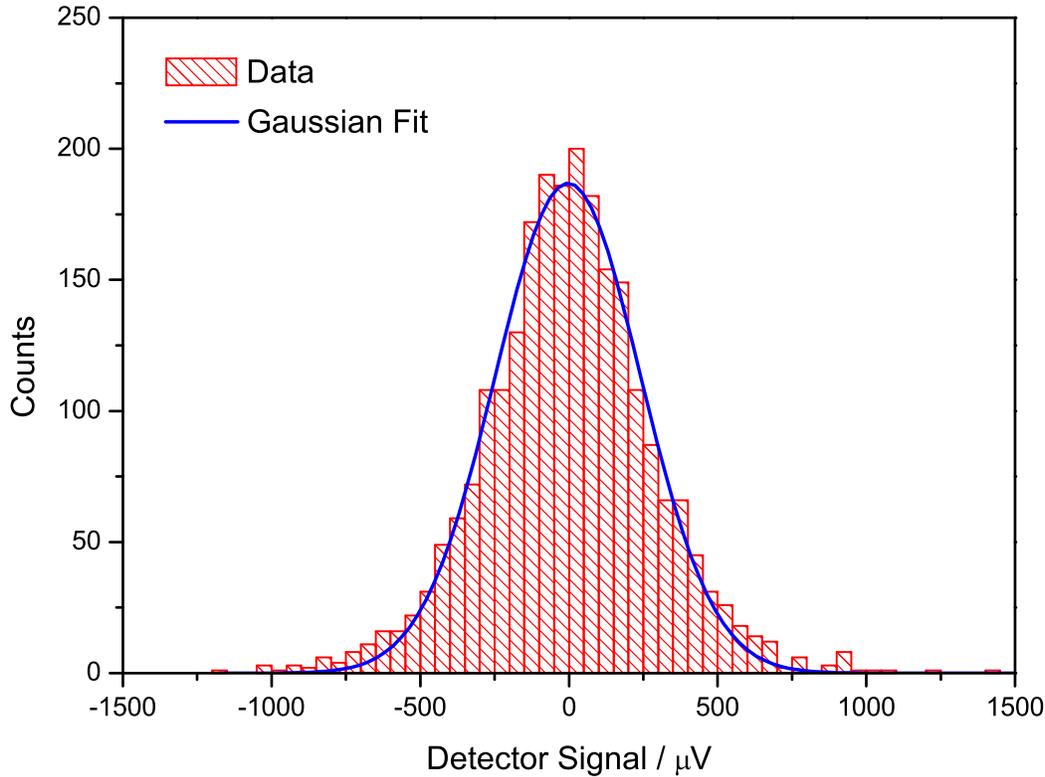}
\caption{\label{fig12} Histogram of null detector signal and Gaussian fit.}
\end{figure}

The automated measurement system allows us to make continuous measurements over many hours (usually during the night) so that the statistics on the final answer can be improved. Figure~\ref{fig12} shows a histogram of all CCC null detector values, expressed in terms of the voltage at the SQUID output, obtained from individual data sets of 10 current reversals. The SQUID output voltage is the correct parameter to combine to give a distribution, since the noise recorded by this detector does not depend on the measurement current in the bridge (the data consists of sets at several different measurement current values). The total number of data sets for the complete measurement campaign was 2300, equating to almost 100 hours of data collection. The solid line is a least-squares Gaussian fit to the histogram and gives a value of $245\ \rm\mu V$ for the standard deviation. Using the SQUID amplifier gain and the sensitivity values in Table~\ref{tab3}, the standard deviation of $245\ \rm\mu V$ translates to a current $I_{ND}$ of 150~fA. Finally, using equation~\ref{eq2} and a measurement current of 100~$\mu$A, this standard deviation is equivalent to a relative deviation of $3.1 \times 10^{-9}$ in the resistance ratio. Since the measurements were done at a range of currents, an effective number of measurements has to be calculated from 

\begin{equation}
\label{eq6}
N_{eff}=\Sigma n_i I_0^2/I_i^2 
\end{equation}

where $I_0 = 100\ \rm\mu A$, and $n_i$ is the number of measurements taken at measurement current $I_i$. This equation gives $n_{eff}=1057$ in our case and dividing the relative uncertainty by the square root of this number leads to a final type~A uncertainty for the entire data set of $9.4 \times 10^{-11}$ compared with $8.6 \times 10^{-11}$ obtained in the next section.

\begin{table}
	\caption{Uncertainty budget.}
	\label{tab4}
	\centering
		\begin{tabular}{ll}
		\br
			 							Contribution			& $(\rm p\Omega/\Omega)$      \\
			\mr
			Reproducibility 	&   		86	  \\
			CCC ratio error  	&       11    \\
			Leakage           &       10    \\
			Servo error       &        6    \\
		  Miscellaneous  	 	&   		$<$ 1	\\
			\br
			Combined standard uncertainty	& 87      \\
			\br
		\end{tabular}
\end{table}

\section{Comparison result}

\begin{figure}
\includegraphics{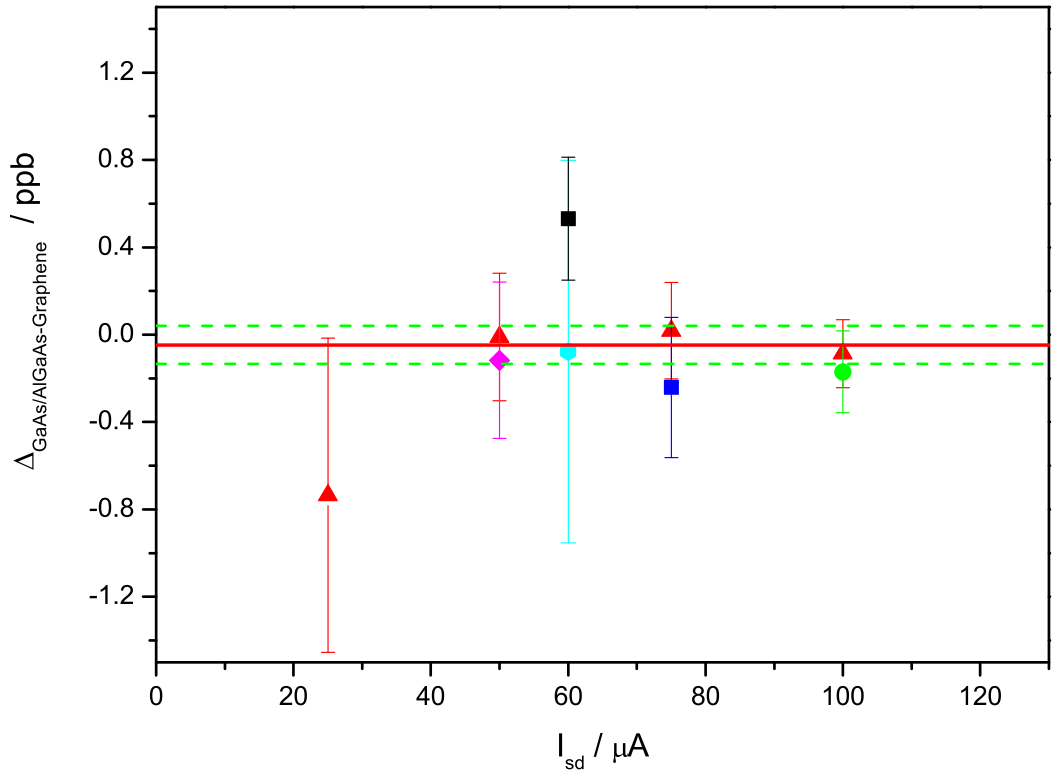}
\caption{\label{fig13} Measurement of $\Delta_{\rm GaAs/AlGaAs-Graphene}$ for the direct comparison of $R_H({\rm GaAs/AlGaAs})$ and $R_H({\rm Graphene})$ as a function of $I_{SD}$. The uncertainty bars represent the $\pm 1\sigma$ standard deviation of the mean. Different symbols are explained in the text. The red line is the weighted mean of all the data points and the green lines signify $\pm 1\sigma$. }
\end{figure}

In order to test the universality of the quantum Hall effect, two different devices need to be set up at the same quantum Hall plateau so that a one-to-one comparison of  resistance can be made (in principle one could also compare different index plateaux if different ratios are available on the CCC). Unfortunately, the quantum Hall plateaux of the graphene and the GaAs/AlGaAs devices do not overlap and two cryostats have to be used for this experiment, one of which was the traveling quantum Hall system of the BIPM. Hartland {\it et al.}~\cite{Hartland1991} used a clever method of tilting one device with respect to the magnetic field direction to make the centers of the plateaux in GaAs and Si occur at the same magnetic field, a trick not currently possible in our quantum Hall probe. As explained in the previous section using two cryostats does lead to a significant increase in noise and subsequent lengthening of the measurement time.

The graphene sample was mounted in a 14~T/300~mK cryostat and connected to the Slave side of the CCC bridge. Two GaAs/AlGaAs samples were mounted in the transportable 11.7~T/1.5~K cryostat provided by the BIPM and connected to the Master side. Prior to making comparison measurements all samples were fully characterised~\cite{Delahaye2003}. For the graphene sample the $B_{\nu=2}$ was set to 14~T, our maximum available magnetic field, and either 9.5~T or 10.5~T for samples PTB2 and LEP, respectively. The winding ratio on the bridge was always 1600:1600 turns. The red triangles in Fig.~\ref{fig13} are the results for GR9 against PTB2 for 4 different source-drain currents in the devices (using contacts 2 and 7 on the graphene device for $R_H$ and contacts 4 and 5 on the GaAs device\footnote{The contact labels for the GaAs devices are the same as for the graphene device depicted in fig.~\ref{fig1}}). The pink diamond is a measurement for GR9 against LEP at at $I_{sd}=50\ \rm\mu A$ (using the same contact configuration as for the red triangle measurements). Here $\Delta_{\rm GaAs/AlGaAs-Graphene}=[R_H({\rm GaAs/AlGaAs})-R_H({\rm Graphene})]/(R_{\rm K}/2)$ and each data point consists of an average of between 3 and 10 hours worth of data. The uncertainty increases for lower $I_{sd}$ because the signal-to-noise ratio is worse for lower $I_{sd}$.

To eliminate the possibility of errors due to non-zero $\rho_{xx}$ we repeated a number of these measurements for non-opposite contacts. Note that it is very difficult to measure $\rho_{xx}$ directly to the required level of precision (see for example fig.~\ref{fig4}). The green dot is a measurement using contacts 3 and 4 for $R_H({\rm GaAs/AlGaAs})$ on the PTB2 device at $I_{sd}=100\ \rm\mu A$. The blue square is a measurement using contacts 2 and 4 for $R_H({\rm Graphene})$ on the graphene device at $I_{sd}=75\ \rm\mu A$. Another test to check for small errors is to reverse the direction of magnetic field on the graphene sample. The results of this measurement is represented by the light blue hexagon for $I_{sd}=60\ \rm\mu A$ (using the same contact configuration as for the red triangle measurements). Finally, the devices were exchanged between the NPL and BIPM cryostats in order to check for small parasitic leakages (the magnetic field was set to 11.5~T for the graphene device in the BIPM cryostat). This data point is shown as the black square in Fig.~\ref{fig13} (using the same contact configuration as for the red triangle measurements). 

The data in Fig.~\ref{fig13} show no sign of any systematic errors in the measurement campaign and so all results can be combined to give a weighted mean of $\Delta_{\rm GaAs/AlGaAs-Graphene}=(-4.7\pm 8.6) \times 10^{-11}$. Combining this with the total type B uncertainty determined in Sec.~\ref{TypeB} gives $8.7\times10^{-11}$ as a final combined uncertainty on the difference between $R_H({\rm GaAs/AlGaAs})$ and $R_H({\rm Graphene})$\footnote{This relative uncertainty is fractionally larger $(0.1\times 10^{-11})$ than the one published in Ref.~\cite{Janssen2011b} due to a more thorough determination of the systematic components.}.

\section{Conclusion}
Previously our knowledge of the universality of the QHE has been limited to the level of 2 or $3\times\ 10^{-10}$ for comparisons between GaAs and Si or between identical GaAs devices~\cite{Hartland1991,Jeckelmann1997,Schopfer2007,Poirier2009}. However both GaAs and Si are traditional semiconductors with a parabolic bandstructure and governed by the same physics. Graphene is a semi metal with a linear bandstructure and is described by Dirac-type massless charge carriers and so universality in terms of material independence goes well beyond the comparison between two semiconductors. It does directly support the Thouless-Laughlin argument~\cite{Thouless1994} that the Hall conductivity is a topological invariant and is a fundamental test of condensed matter theory. 

Our results on material independence is the strongest evidence yet that the hypothesis that the resistance is quantised in units of $h/e^2$ is correct and thereby supports the pending redefinition of the SI units kilogram and ampere in terms of $h$ and $e$. Note that the correctness of this equation can only be shown in a comparison of $R_{\rm H}$ with an independent realisation of $h/e^2$, for example via the Thompson-Lampard capacitance~\cite{Bachmair2009}. However, the relative uncertainty which can be obtained in such an experiment is some 3 orders of magnitude worse than in a test of the material independence. 

In our universality experiment the maximum source-drain current that the GaAs device can sustain without dissipation limits the measurement uncertainty, whereas a potentially lower uncertainty can be obtained in a consistency check of two graphene devices. Also the measurement system could still be significantly improved by reducing the excess noise or by adopting a different measurement technique such as demonstrated in Ref.~\cite{Schopfer2007}.

The journey from the original discovery of the QHE in graphene to a quantum resistance standard which outperforms the established technology in many aspects has been remarkably short. For epitaxial graphene the robustness of the quantisation in terms of temperature, magnetic field and source-drain current is exceptional. The material is cheap and relatively easy to fabricate and process. It allows for the realisation of a quantum resistance standard with modest means, e.g. a small superconducting magnet and cryocooler. As such it will improve the proliferation of quantum standards and allow many smaller laboratories to realise their own resistance scale. One even could envisage university students being able to perform QHE experiments, much in the same way as the discovery of high-$T_c$ superconductors enabled table-top experiments with Josephson junctions in many science classes.

\ack
The authors are grateful to Dale Henderson, Stephen Giblin, Pravin Patel for their valuable contributions during this work.  This work was supported by the NMS Pathfinder Programme, Swedish Research Council and Foundation for Strategic Research, EU FP7 STREPs ConceptGraphene and SINGLE, EPSRC grant EP/G041954 and the Science \& Innovation Award EP/G014787.


\section*{References}



\begin{thebibliography}{10}

\bibitem{Novoselov2005}
K.~S. Novoselov, A.~K. Geim, S.~V. Morozov, D.~Jiang, M.~I. Katsnelson, I.~V.
  Grigorieva, S.~V. Dubonos, and A.~A. Firsov.
\newblock Two-dimensional gas of massless dirac fermions in graphene.
\newblock {\em Nature}, {\bf 438} 197-200, 2005.

\bibitem{Zhang2005}
Y.~B. Zhang, Y.~W. Tan, H.~L. Stormer, and P.~Kim.
\newblock Experimental observation of the quantum hall effect and berry's phase
  in graphene.
\newblock {\em Nature}, {\bf 438} 201-204, 2005.

\bibitem{Prange1990}
For a review, see R.~E. Prange and S.~M. Girvin.
\newblock {\em The Quantum Hall Effect}.
\newblock Springer-Verlag, Berlin, 1990.

\bibitem{Jeckelmann2001}
B.~Jeckelmann and B.~Jeanneret.
\newblock The quantum hall effect as an electrical resistance standard.
\newblock {\em Reports on Progress in Physics}, {\b 64} 1603-1655, 2001.

\bibitem{Geim2009}
A.~K. Geim.
\newblock Graphene: Status and prospects.
\newblock {\em Science}, {\bf 324} 1530-1534, 2009.

\bibitem{Novoselov2007}
K.~S. Novoselov, Z.~Jiang, Y.~Zhang, S.~V. Morozov, H.~L. Stormer, U.~Zeitler,
  J.~C. Maan, G.~S. Boebinger, P.~Kim, and A.~K. Geim.
\newblock Room-temperature quantum hall effect in graphene.
\newblock {\em Science}, {\bf 315} 1379-1379, 2007.

\bibitem{Thouless1994}
D.~J. Thouless.
\newblock Topological interpretations of quantum hall conductance.
\newblock {\em Journal of Mathematical Physics}, {\bf 35} 5362-5372, 1994.

\bibitem{Hehl2004}
F.~W. Hehl, Y.~N. Obukhov, and B.~Rosenow.
\newblock Is the quantum hall effect influenced by the gravitational field?
\newblock {\em Physical Review Letters}, {\bf 93} 096804-4 2004.

\bibitem{Penin2009}
A.~A. Penin.
\newblock Quantum hall effect in quantum electrodynamics.
\newblock {\em Physical Review B}, {\bf 79} 113303-7, 2009.

\bibitem{Mohr2008b}
P.~J. Mohr, B.~N. Taylor, and D.~B. Newell.
\newblock Codata recommended values of the fundamental physical constants:
  2006.
\newblock {\em Reviews of Modern Physics}, {\bf 80} 633-730, 2008.

\bibitem{Mills2011}
I.~M. Mills, P.~J. Mohr, T.~J. Quinn, B.~N. Taylor, and E.~R. Williams.
\newblock Adapting the international system of units to the twenty-first
  century.
\newblock {\em Philosophical Transactions of the Royal Society and Mathematical
  Physical and Engineering Sciences}, {\bf 369} 3907-3924, 2011.

\bibitem{Delahaye2003}
F.~Delahaye and B.~Jeckelmann.
\newblock Revised technical guidelines for reliable dc measurements of the
  quantized hall resistance.
\newblock {\em Metrologia}, {\bf 40} 217-223, 2003.

\bibitem{Giesbers2008}
A.~J.~M. Giesbers, G.~Rietveld, E.~Houtzager, U.~Zeitler, R.~Yang, K.~S.
  Novoselov, A.~K. Geim, and J.~C. Maan.
\newblock Quantum resistance metrology in graphene.
\newblock {\em Applied Physics Letters}, {\bf 93} 222109-12, 2008.

\bibitem{Shen2009}
T.~Shen, J.~J. Gu, M.~Xu, Y.~Q. Wu, M.~L. Bolen, M.~A. Capano, L.~W. Engel, and
  P.~D. Ye.
\newblock Observation of quantum-hall effect in gated epitaxial graphene grown
  on sic (0001).
\newblock {\em Applied Physics Letters}, {\bf 95} 172105-3, 2009.

\bibitem{Wu2009}
X.~S. Wu, Y.~K. Hu, M.~Ruan, N.~K.~Madiomanana, J.~Hankinson, M.~Sprinkle,
  C.~Berger, and W.~A.~de~Heer.
\newblock Half integer quantum hall effect in high mobility single layer
  epitaxial graphene.
\newblock {\em Applied Physics Letters}, {\bf 95} 223108-3, 2009.

\bibitem{Tzalenchuk2010}
A.~Tzalenchuk, S.~Lara-Avila, A.~Kalaboukhov, S.~Paolillo, M.~Syvajarvi,
  R.~Yakimova, O.~Kazakova, T.~J.~B.~M.~Janssen, V.~Fal'ko, and S.~Kubatkin.
\newblock Towards a quantum resistance standard based on epitaxial graphene.
\newblock {\em Nature Nanotechnology}, {\bf 5} 186-189, 2010.

\bibitem{Jobst2010}
J.~Jobst, D.~Waldmann, F.~Speck, R.~Hirner, D.~K.~Maude, T.~Seyller, and 
H.~B.~Weber.
\newblock Quantum oscillations and quantum hall effect in epitaxial graphene.
\newblock {\em Physical Review B}, {\bf 81} 195434-6, 2010.

\bibitem{Tanabe2010}
S.~Tanabe, Y.~Sekine, H.~Kageshima, M.~Nagase, and H.~Hibino.
\newblock Half-integer quantum hall effect in gate-controlled epitaxial
  graphene devices.
\newblock {\em Applied Physics Express}, {\bf 3} 075102-3, 2010.

\bibitem{Williams2010}
J.~M.~Williams, T.~J.~B.~M.~Janssen, G.~Rietveld, and E.~Houtzager.
\newblock An automated cryogenic current comparator resistance ratio bridge for
  routine resistance measurements.
\newblock {\em Metrologia}, {\bf 47} 167-174, 2010.

\bibitem{Janssen2011b}
T.~J.~B.~M.~Janssen, N.~E.~Fletcher, R.~Goebel, J.~M.~Williams, A.~Tzalenchuk,
  R.~Yakimova, S.~Lara-Avila, S.~Kubatkin, and V.~I.~Fal'ko.
\newblock Graphene, universality of the quantum hall effect and redefinition of
  the SI system.
\newblock {\em New Journal of Physics}, {\bf 13} 093026-6, 2011.

\bibitem{Woszczyna2011}
M.~Woszczyna, M.~Friedemann, K.~Pierz, T.~Weimann, and F.~J.~Ahlers.
\newblock Magneto-transport properties of exfoliated graphene on GaAs.
\newblock {\em Journal of Applied Physics}, {\bf 110} 043712-6, 2011.

\bibitem{Guignard2011}
J.~Guignard, F.~Schopfer, W.~Poirier, and D.~C.~Glattli.
\newblock Quantum Hall effect in exfoliated graphene on Si/SiO$_2$ from metrological measurements.
\newblock {\em arXiv:cond-mat/1110.4884v1}.

\bibitem{Virojanadara2008}
C.~Virojanadara, C~Virojanadara, M.~Syvajarvi, R.~Yakimova, L.~I.~Johansson,
  A.~A.~Zakharov, and T.~Balasubramanian.
\newblock Homogeneous large-area graphene layer growth on 6h-sic(0001).
\newblock {\em Physical Review B}, {\bf 78} 245403-6, 2008.

\bibitem{Kopylov2010}
S.~Kopylov, A.~Tzalenchuk, S.~Kubatkin, and V.~I. Fal'ko.
\newblock Charge transfer between epitaxial graphene and silicon carbide.
\newblock {\em Applied Physics Letters}, {\bf 97} 112109-3, 2010.

\bibitem{Lara-Avila2011}
S.~Lara-Avila, K.~Moth-Poulsen, R.~Yakimova, T.~Bjornholm, V.~Fal'ko,
  A.~Tzalenchuk, and S.~Kubatkin.
\newblock Non-volatile photochemical gating of an epitaxial graphene/polymer
  heterostructure.
\newblock {\em Advanced Materials}, {\bf 23} 878-882, 2011.

\bibitem{Lara-Avila2011b}
S.~Lara-Avila, A.~Tzalenchuk, S.~Kubatkin, R.~Yakimova, T.~J.~B.~M.~Janssen, K.~Cedergren, T.~Bergsten, and V.~Fal'ko.
\newblock Disordered Fermi liquid in epitaxial graphene from quantum transport measurements.
\newblock {\em Physical Review Letters}, {\bf 107} 166602-5, 2011.

\bibitem{Yoshioka2002}
D. Yoshioka, {\em The quantum Hall effect} (Springer-Verlag, Berlin 2002).

\bibitem{Cage1983}
M.~E.~Cage, R.F.~Dziuba, B.~F.~Field, E.~R.~Williams, S.~M.~Girvin, A.~C.~Gossard, D.~C.~Tsui, and R.~J.~Wagner.
\newblock Dissipation and dynamic nonlinear behavior in the quantum Hall regime.
\newblock {\em Physical Review Letters}, {\bf 51} 1374, 1983.

\bibitem{vanderwel1988}
W.~van der Wel, C.~J.~P.~M.~Harmans, J.~E.~Mooij.
\newblock A geometric explanation of the temperature dependence of the quantised Hall effect.
\newblock {\em Journal of Physics C: Solid State Physics}, {\bf 21} L171-L175, 1988.

\bibitem{Janssen2011a}
T.~J.~B.~M.~Janssen, A.~Tzalenchuk, R.~Yakimova, S.~Kubatkin, S.~Lara-Avila,
  S.~Kopylov, and V.~I. Fal'ko.
\newblock Anomalously strong pinning of the filling factor $\nu=2$ in epitaxial
  graphene.
\newblock {\em Physical Review B}, {\bf 83} 233402-4, 2011.

\bibitem{Williams2011}
J.~M.~Williams, G.~Rietveld, E.~Houtzager, and T.~J.~B.~M.~Janssen.
\newblock Design considerations for a ccc bridge with complete digital control.
\newblock {\em IEEE Transactions on Instrumentation and Measurement}, {\bf 99} 1-6, 2011.

\bibitem{Delahaye1991}
F.~Delahaye and D.~Bournaud.
\newblock Low-noise measurements of the quantized hall resistance using an
  improved cryogenic current comparator bridge.
\newblock {\em IEEE Transactions on Instrumentation and Measurement},
  {\bf 40} 237-240, 1991.

\bibitem{Williams1991}
J.~M.~Williams and A.~Hartland.
\newblock An automated cryogenic current comparator resistance ratio bridge.
\newblock {\em IEEE Transactions on Instrumentation and Measurement},
  {\bf 40} 267-270, 1991.

\bibitem{Witt2000}
T.~J.~Witt and D.~Reymann.
\newblock Using power spectra and allan variances to characterise the noise of
  zener-diode voltage standards.
\newblock {\em IEE Proceedings-Science Measurement and Technology},
  {\bf 147} 177-182, 2000.

\bibitem{Hartland1991}
A.~Hartland, K.~Jones, J.~M.~Williams, B.~L.~Gallagher, and T.~Galloway.
\newblock Direct comparison of the quantized hall resistance in
  gallium-arsenide and silicon.
\newblock {\em Physical Review Letters}, {\bf 66} 969-973, 1991.

\bibitem{Jeckelmann1997}
B.~Jeckelmann and B.~Jeanneret.
\newblock High-precision measurements of the quantized hall resistance:
  Experimental conditions for universality.
\newblock {\em Physical Review B}, {\bf 55} 13124-13134, 1997.

\bibitem{Schopfer2007}
F.~Schopfer and W.~Poirier.
\newblock Testing universality of the quantum hall effect by means of the
  wheatstone bridge.
\newblock {\em Journal of Applied Physics}, {\bf 102} 054903-9, 2007.

\bibitem{Poirier2009}
W.~Poirier and F.~Schopfer.
\newblock Resistance metrology based on the quantum hall effect.
\newblock {\em European Physical Journal-Special Topics}, {\bf 172} 207-245, 2009.

\bibitem{Bachmair2009}
H.~Bachmair.
\newblock Determination of the unit of resistance and the von Klitzing constant
$R_{\rm K}$ based on a calculable capacitor.
\newblock {\em European Physical Journal-Special Topics}, {\bf 172} 257-266, 2009.

\end{thebibliography}

\end{document}